\documentclass[11pt, fleqn]{article}
\usepackage[utf8]{inputenc}
\usepackage{amsmath,amssymb}
\usepackage{graphicx}
\usepackage{subfig}
\usepackage{here}
\usepackage{float}
\usepackage{multirow}
\usepackage[margin=0.75in]{geometry}
\usepackage{boldline}
\usepackage{xcolor}
\usepackage{ulem}
\usepackage{authblk}
\usepackage{siunitx}
\usepackage{booktabs}
\usepackage{caption}
\usepackage{comment}
\usepackage[percent]{overpic}
\usepackage{mathtools}
\usepackage{hyperref}
\usepackage[table]{xcolor}
\usepackage{cite,physics}

\newcommand{\eg}{\textit{e.g.,~}}
\newcommand{\ie}{\textit{i.e.,~}}

\newcommand{\gev}{\ensuremath{\,\mathrm{GeV}}}
\newcommand{\aub}{a_\tau^\uparrow}
\newcommand{\alb}{a_\tau^\downarrow}

\newcommand{\ga}{\left(}
\newcommand{\dr}{\right)}

\DeclareMathAlphabet{\mathpzc}{OT1}{pzc}{m}{it}

\newcommand{\rom}[1]{\uppercase\expandafter{\romannumeral #1\relax}}
\def\au{a_\mu}
\def\at{a_\tau}
\def\atQCD{a_\tau^{\mathrm{QCD}}}

\def\auhvplo{a_\mu^\mathrm{HVP,\,LO}}

\def\athvplo{a_\tau^\mathrm{HVP,\,LO}}

\title{Bounds on hadronic contributions to the anomalous magnetic moment of the $\tau$ lepton from QCD Sum Rules}
\author[1]{Siyuan\ Li\thanks{Siyuan.Li@uni-mainz.de}}
\author[2]{J.\ Ho\thanks{Jason.Ho@dordt.edu}}
\author[3,4]{R.T.\ Kleiv\thanks{Robin.Kleiv@tru.ca}}
\author[3]{D.\ Leighton\thanks{Daylen.Leighton@outlook.com}}
\author[2,5]{E.J.\ Obbink\thanks{Eobbink@iu.edu}}
\author[4]{T.G.\ Steele\thanks{Tom.Steele@usask.ca}}
\affil[1]{Institut f\"ur Kernphysik,
 Johannes Gutenberg-Universit\"at  Mainz,  D-55128 Mainz, Germany}
\affil[2]{Department of Physics\\ Dordt University\\ Sioux Center, IA, 51250, USA}
\affil[3]{Department of Physical Sciences (Physics)\\ Thompson Rivers University\\ Kamloops, BC, V2C~0C8, Canada}
\affil[4]{Department of Physics and Engineering Physics\\ University of Saskatchewan\\ Saskatoon, SK, S7N 5E2, Canada}
\affil[5]{Department of Physics, Indiana University, Bloomington, IN, 47405 USA}

\begin{document}
\maketitle
\begin{abstract}
A QCD sum-rule study is performed to place bounds on the leading-order (LO) hadronic vacuum polarization (HVP) contribution to the anomalous magnetic moment of the $\tau$ lepton ($\athvplo,\, \at = (g-2)_\tau/2$). The QCD contributions to $\athvplo$ are modeled as a combination of Laplace QCD sum rules (LSRs) and finite-energy QCD sum rules (FESRs) of various weights, which are bounded by 
H\"older and related fundamental inequalities.
The light-quark ($u,\,d,\,s$) contributions are evaluated up to five-loop order in perturbative contributions in the chiral limit, LO in light quark mass corrections, next-to-leading order (NLO) in dimension-four QCD condensates and to LO in dimension-six QCD condensates. Extending previous work on $a_\mu^\mathrm{HVP,LO}$, 
we present a robust methodology for using QCD sum rules to constrain $\athvplo$, 
and more generally any lepton $g-2$. 
A criterion is developed to extract from the bounds analysis a final result of $\at^\mathrm{QCD} = \ga 3.385\pm 0.048 \dr \times 10^{-6}$, which after inclusion of charmonium and bottomonium contributions gives a final result  $\athvplo = \ga 3.501\pm0.048\dr \times 10^{-6}$, comparable to data-driven and lattice predictions. 
Lattice QCD results along with our value for $\at^\mathrm{HVP,LO}$ are used to  obtain a combined QCD determination from lattice and sum-rule methods 
of 
$\auhvplo = (7.217\pm0.233)\times10^{-8}$. 
Our analysis provides fundamental QCD sum-rule inputs into the theoretical determinations of $\at$, and establishes new methodologies for QCD sum-rule determinations of low-energy observables. 
\end{abstract}

\section{Introduction}
Motivated by the potential contribution from physics beyond the Standard Model (BSM), the anomalous magnetic moments of charged leptons ($a_\ell=(g-2)_\ell/2,\, l\in \{e,\mu,\tau\}$)~\cite{Keshavarzi:2019abf,DiLuzio:2024sps,Giusti:2019hkz,Budapest-Marseille-Wuppertal:2017okr,Giusti:2020efo,Jegerlehner:2017zsb,Hoferichter:2025fea,Fan:2022eto,Benmerrouche:1993rj,Holdom:1993ad,Ahmady:2002pa,Hamzeh:1996np,Narison:2001jt,Eidelman:2007sb,PhysRevD.102.055022}, have been studied extensively in both experiment and theory. Among them, the determination of the muon anomalous magnetic moment has driven significant advances in both experimental measurement techniques and theoretical methods for its phenomenological data-driven and lattice QCD (LQCD) evaluation (see \eg reviews Refs.~\cite{Aguillard:2025fij,Aliberti:2025beg,Boccaletti:2024guq}).

In our previous work, we developed fundamental upper and lower bounds on the leading-order (LO) hadronic vacuum polarization (HVP) contribution to the muon $\left(g-2\right)$, $a_\mu^\mathrm{HVP,LO}$, using the H\"older Inequality and finite-energy QCD sum rules (FESRs)~\cite{Li:2024frm}. 
Our result is consistent with both the data-driven dispersive and LQCD determinations of $\au^\mathrm{HVP,LO}$, despite the persistent tension between these two approaches and within different data-driven determinations.

In light of recent agreement between LQCD and experimental measurements of $\au$~\cite{Aliberti:2025beg,Boccaletti:2024guq}, we proceed to provide QCD Laplace sum rule (LSR) bounds on HVP contribution to the $\tau$ lepton. As a heavier lepton than $\mu$, the tau lepton has a higher sensitivity to potential new particles from BSM physics, amplified by a factor of $m_\tau^2/m_\mu^2\simeq 283$. 
The HVP contribution to $\at$ with the same strong interaction contributions are similarly amplified relative to $\au$, as well as for the possible BSM contributions. 

The challenge in measuring $a_\tau$ lies in the short lifetime of the $\tau$, which is $2.90\,\times10^{-13}\,\mathrm{s}$~\cite{ParticleDataGroup:2026aaa}, rendering the storage ring approach used in the $\au$ measurement at Fermilab infeasible~\cite{Aguillard:2025fij,Aliberti:2025beg}. 
Indirect methods using $\gamma\gamma\rightarrow\tau\tau$ from processes such as high-energy proton-proton collisions by the CMS collaboration at the LHC brought improved limits of $\at = 0.001(3)$~\cite{CMS:2024qjo}, compared to the previous results from nuclei collisions by ATLAS~\cite{ATLAS:2022ryk} and CMS~\cite{CMS:2022arf}, and from $e^+e^-$ collisions by the DELPHI collaboration~\cite{DELPHI:2003nah}. On the theoretical side, $\at$ has also been evaluated through LQCD (see \eg Ref.~\cite{Budapest-Marseille-Wuppertal:2017okr}) and the data-driven approaches (see \eg Refs.~\cite{Eidelman:2007sb,PhysRevD.102.055022}), with $\athvplo =(332.81\pm1.39)\times10^{-8}$ determined by the KNT collaboration~\cite{Keshavarzi:2019abf}.

In this work, we derive a set of bounds on $\athvplo$ by combining FESRs and LSRs, obtained by fitting the kernel function of the $a_\tau$ dispersion relation.
The QCD-hadron duality underlying FESRs~\cite{Floratos:1978jb,Hubschmid:1980rm,Bertlmann:1984ih,Caprini:1985ex} and LSRs~\cite{Shifman:1978by,Shifman:1978bx} allows integrals over the hadronic spectral function to be related to QCD predictions of strong-interaction observables~\cite{Shifman:1978bx,Shifman:1978by} (see \eg Refs.~\cite{Reinders:1984sr,Narison:2002woh,Gubler:2018ctz,Colangelo:2000dp} for QCD sum rule reviews). In the high-energy limit, the resulting FESRs have a structure that can be related to the dispersion relation for $a_\ell$, allowing sum-rule techniques to be applied to the study of leptons, as demonstrated for the muon in Ref.~\cite{Li:2024frm}. Extending this idea by expressing $\at$ in terms of polynomial combinations of QCD sum rules, we obtain upper and lower bounds on the observable using H\"older's inequality~\cite{Beckenbach:1961,Berberian:1965}, together with the corresponding subsidiary conditions discussed in Sec.~\ref{sec:kernel_fit}. The explicit sum-rule expressions are provided in the Appendix. The QCD inputs used in our bound expressions for light quark contributions are included in Sec.~\ref{sec:QCD_input}. The perturbative contribution is provided up to five-loop order in the $\overline{\mathrm{MS}}$-scheme, in the chiral limit, and LO in quark mass corrections. Non-perturbative contributions include next-to-leading order (NLO) dimension-four QCD condensates and LO in dimension-six QCD condensates.
Section~\ref{sec:analysis} presents the analysis and final results. A summary is provided in Sec.~\ref{sec:summary}.

\section{Kernel Function Fitting and QCD Sum-Rule Bounds}
\label{sec:kernel_fit}
\begin{figure}
    \centering
    \includegraphics[width=0.55\linewidth]{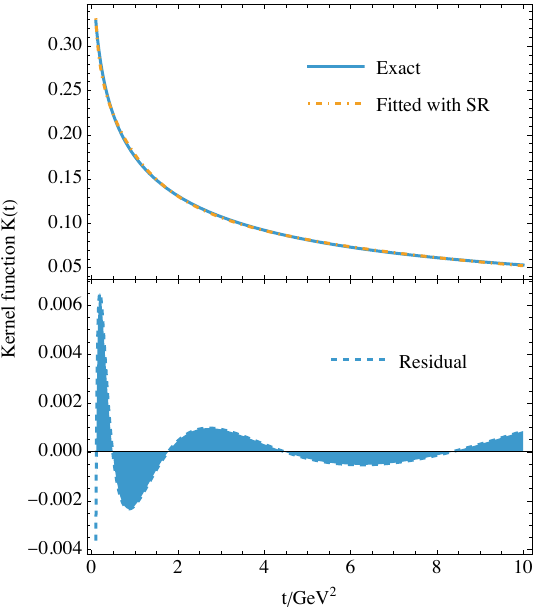}
    \caption{Comparison and residual (exact minus fit) plots of kernel function fitting with the polynomial combination of LSRs and FESRs shown in Eq.~\eqref{eq.general-fit} and using fitting parameters in Table~\ref{tab:fitting_paras}. The source of error associated with the residuals is negligible compared to QCD parameter uncertainties (see Sec.~\ref{sec:analysis}).}
    \label{fig:kernel_fit}
\end{figure}

The dispersion relation for $\athvplo$ takes the same form as for all leptons~\cite{Aliberti:2025beg,Steele1991,Lautrup:1974ic}
\begin{equation}\label{eq:a_tau_0}
    a_\tau^{\mathrm{HVP,\,LO}} = \frac{1}{4\pi^3} \int_{t_0}^\infty \sigma^H(t) K_\tau(t)\,\mathrm{d}t \,,
\end{equation}
where $\sigma^H$ is the hadronic cross section and the kernel function is
\begin{equation}
    K_\tau(t) =  \int_{0}^{1} \,dy \frac{y^2\left(1-y\right)}{y^2 + \left(1-y\right)t/m_\tau^2} .
    \label{eq:exact-kernel}
\end{equation}
We use the hadronic $R$-ratio
\begin{gather}
    R(t) = \frac{\sigma^H(t)}{\sigma\left(e^+e^-\rightarrow\mu^+\mu^-\right)} \,, 
    \label{eq:RtIm}
\end{gather}
 and the muon production cross section
\begin{gather}
    \sigma\left(e^+e^-\rightarrow\mu^+\mu^-\right) = \frac{4\pi\alpha^2}{3t^2}\left(t+2m_\mu^2\right)\sqrt{1-\frac{4m_\mu^2}{t}} 
    = \frac{4\pi \alpha^2}{3t} + \mathcal{O}\left(\frac{1}{t^3}\right).
    \label{eq:epemsigma}
\end{gather}
Using Eq.~\eqref{eq:epemsigma} in Eq.~\eqref{eq:RtIm}, expressing 
the hadronic vacuum polarization spectral function as $R(t) = 12\pi \mathrm{Im}\Pi^H(t)$ \cite{Caprini:1984ud,Steele1991}, isolating $\sigma^H(t)$, and substituting the resulting expression into Eq.~\eqref{eq:a_tau_0}, we find
\begin{equation} \label{eq:a_tau_QCD}
   a_\tau^{\mathrm{HVP,LO}} =  4\alpha^2 \int_{t_0}^\infty \frac{1}{\pi}\mathrm{Im}\Pi^H(t)\frac{1}{t}K_\tau(t)\,\mathrm{d}t,
\end{equation}
where 
$\frac{1}{\pi}\mathrm{Im}\Pi^H(t)$ is the hadronic spectral function. 
From the QCD perspective, the hadronic spectral function below the charm threshold   is related to the light-quark 
$(u,d,s)$
vector-current correlation function $\Pi^H= \sum_{q} Q_q^2 \Pi_q$ with quark charge pre-factors $Q_q^2$ 
\begin{equation}
   \Pi^{\mu\nu}_q=
\left(p^\mu p^\nu-p^2g^{\mu\nu} \right) \Pi_q\left(p^2\right)
=i\int d^4x \,e^{ip\cdot x}
\left\langle O \right\vert 
T\left[j^\mu(x)j^\nu(0)\right]\left\vert O\right\rangle
\,, \quad j^\mu = \bar{q}\gamma^\mu q \,.
\label{tom_corr_fn}
\end{equation}
Hereafter we use the notation $\athvplo$ to denote the full hadronic contribution (including the effects of heavy flavors, \eg charmonium and bottomonium states), and the notation $\at^\mathrm{QCD}$ to denote the light-quark QCD prediction within $\athvplo$.
In the HVP contribution, the lower limit of the integral in Eq.~\eqref{eq:a_tau_0} is related to the pion-production threshold $t_0=(2 m_\pi)^2$. In our previous work~\cite{Li:2024frm}, we approximated $K_\mu(t)$ by assuming near-massless final lepton states, \ie $t\gg m_\mu^2$, leading to an  $\au$ expression as a negative-weight FESR. With the much larger $\tau$ lepton mass, we instead fit $K_\tau(t)$ to the following form in the crucial low-energy region below the charm threshold:
\begin{equation}
    K_\mathrm{fit}(t) = \sum_n A_n t^n + \sum_i e^{-t \tau_i} \ga\sum_m B_{i,m} t^m \dr,
    \label{eq.general-fit}
\end{equation}
where $\{A_n,B_m,\tau_i\}$ are constant fitting coefficients with values shown in Table~\ref{tab:fitting_paras}.
We note that fitted values of $\tau_i$ are safely in a reliable region for LSRs of light-quark vector currents. The fitted kernel function~\eqref{eq.general-fit} is compared to the exact form in Eq.~\eqref{eq:exact-kernel}, which is shown in Fig.~\ref{fig:kernel_fit}. 
Such a fitting structure can be written in terms of FESRs and LSRs after inserting the fitted kernel function back into Eq.~\eqref{eq:a_tau_QCD} for $a_\tau^{\mathrm{QCD}}$:
\begin{equation}\label{eq:a_fit_exp}
    a_\tau^{\mathrm{QCD}} 
    \approx  4\alpha^2 \int_{t_0}^\infty \frac{1}{\pi}\mathrm{Im}\Pi^H(t)\frac{1}{t}K_\mathrm{fit}(t)\,\mathrm{d}t 
    =\lim_{s_0\rightarrow \infty}4\alpha^2\left(\sum_{n=-1}^{0} A_n F_{n-1}\left(s_0\right) + \sum_{i=1}^{2} \sum_{m=0}^{3} B_{i,m} \mathcal{L}_{m-1}\left(\tau_i,s_0\right)\right),
\end{equation}
thus allowing QCD sum rules to be used to determine $\atQCD$. We discuss subtleties of the implementation of the large $s_0$ limit below. The FESRs and LSRs are defined as
\begin{gather}
    F_k\left(s_0\right) = \int_{t_0}^{s_0} \frac{1}{\pi}\mathrm{Im}\Pi^H(t)\, t^k\,\mathrm{d}t,\label{eq:Fk_def}
    \\
    \mathcal{L}_k\left(\tau_i,s_0\right) = \int_{t_0}^{s_0} \frac{1}{\pi}\mathrm{Im}\Pi^H(t)\, t^k e^{-t \tau_i}\,\mathrm{d}t,\label{eq:Lk_def}
\end{gather}
where $k$ is the weight of the sum rules.
In Eqs.~\eqref{eq:Fk_def}--\eqref{eq:Lk_def} it is implicit that the LHS represents the QCD prediction and the RHS contains the hadronic spectral function $\frac{1}{\pi}\mathrm{Im}\Pi^H(t)$.
For positive weight  $k\ge 0$ sum rules, $s_0$ represents the  energy-scale boundary that separates the resonance and continuum sectors of the spectral function~\cite{Shifman:1978by,Shifman:1978bx,Narison:2002woh}.
However, this interpretation of the QCD continuum $s_0$  is obscured for the inequality bounds in the cases where $k<0$, preventing a naive implementation of a large $s_0$ limit in Eq.~\eqref{eq:a_fit_exp}. In the bounds analysis one needs to carefully scan from high to low energies to identify the appropriate $s_0$. This is discussed further in Sec.~\ref{sec:analysis}.

\begin{table}[!ht]
\centering
\footnotesize
\renewcommand{\arraystretch}{1.4}
\begin{tabular}{c c cc cccc}
\hline\hline
$i$ & $\tau_i$ $\ga\gev^{-2}\dr$ &
\multicolumn{2}{c}{$A_n$ coefficients $\ga\gev^{-2n}\dr$} &
\multicolumn{4}{c}{$B_{i,m}$ coefficients $\ga\gev^{-2m}\dr$} \\
\cline{3-4}\cline{5-8}
 & & $A_{-1}$ & $A_0$ & $B_{i,0}$ & $B_{i,1}$ & $B_{i,2}$ & $B_{i,3}$ \\
\hline
-- & -- &
$2.42\times10^{-2}$ & $8.14\times10^{-3}$ &
-- & -- & -- & -- \\

1 & $2.98\times10^{-1}$ &
-- & -- &
$1.24\times10^{-1}$ & $-1.49\times10^{-2}$ &
$5.07\times10^{-3}$ & $-2.74\times10^{-11}$ \\

2 & $6.14\times10^{-1}$ &
-- & -- &
$1.15\times10^{-1}$ & $-1.51\times10^{-2}$ &
$1.22\times10^{-2}$ & $-1.29\times10^{-12}$ \\
\hline\hline
\end{tabular}

\caption{Fitting coefficients in $K_{\mathrm{fit}}(t)$, defined in
Eq.~\eqref{eq.general-fit}. The coefficients are obtained by approximating
the kernel function over the range
$t\in[0.1,10]~\mathrm{GeV}^{2}$, and the resulting fit is shown in Fig.~\ref{fig:kernel_fit}.
}
\label{tab:fitting_paras}
\end{table}

All sum rules are well-defined and direct predictions can be found for $k\geq 0$. 
From the fitting coefficients listed in Table~\ref{tab:fitting_paras}, one may notice that our kernel fitting extends to negative weight sum rules as low as $k=-2$. Because negative-weight sum rules introduce low-energy constants which are typically unknown, they cannot be evaluated directly (see \eg Refs.~\cite{Caprini:1984ud,Casas:1985yw,Steele1991,Narison:2023srj}). However, constraints can be found for those $\mathcal{L}_k$ and $F_k$ $(k<0)$ (and thus $\at^\mathrm{QCD}$) through fundamental  H\"older inequalities, positivity of the hadronic spectral function, and other related inequality relations (see \eg Refs.~\cite{Benmerrouche:1995qa,Ho:2018cat,Steele:1998ry,Shi:1999hm,Wang:2016sdt,Yuan:2017foa,Dalfovo:1992dr}). 
In Ref.~\cite{Li:2024frm}, we derived lower and upper limits for $\auhvplo$ which involved bounding $F_{-2}(s_0)$ and $F_{-1} (s_0)$.
In obtaining the   FESR bounds in \cite{Li:2024frm} the  positivity of the hadronic spectral function is used to define an integration measure $[d\mu]=\frac{1}{\pi}\mathrm{Im}\Pi^H(t)$. 
Those bounds can be extended from the FESRs to the LSRs via the same techniques as \cite{Li:2024frm} with the integration measure $[d\mu]=\frac{1}{\pi}e^{-t\tau}\mathrm{Im}\Pi^H(t)$:
\begin{subequations}\label{eqs:fm1}
    \begin{align}
        \frac{\mathcal{S}_0^2}{\mathcal{S}_1} & \equiv \mathcal{S}_{-1}^\downarrow 
        \leq \mathcal{S}_{-1}\le \mathcal{S}_{-1}^\uparrow \equiv \mathcal{S}_{-1}^{(B)}=\frac{\mathcal{S}_0}{t_0}-\frac{\left(\mathcal{S}_1/t_0-\mathcal{S}_0\right)^2}{\left(\mathcal{S}_2/t_0-\mathcal{S}_1\right)}\,,
    \label{fm1_B}\\
        \mathcal{S}_2&/t_0 -\mathcal{S}_1 >0\,;
    \label{fm1_B_condition}
    \end{align}
\end{subequations}
\begin{subequations}\label{eqs:fm2}
    \begin{align}
\frac{\mathcal{S}_{0}^3\left(s_0\right)}{\mathcal{S}_1^2\left(s_0\right)} & \equiv
\mathcal{S}_{-2}^\downarrow\le 
\mathcal{S}_{-2} 
\le \mathcal{S}_{-2}^\uparrow \equiv
\left\{
\begin{aligned}
&\mathcal{S}_{-1}^{(B)}/t_0
-\frac{\left(\mathcal{S}_0/t_0-\mathcal{S}_{-1}^{(B)}\right)^2}
{\mathcal{S}_1/t_0-\mathcal{S}_0}
\\[-2pt]
&\mathcal{S}_0/t_0^2
-\frac{\left(\mathcal{S}_1/t_0^2-\mathcal{S}_{-1}^{(B)}\right)^2}
{\mathcal{S}_2/t_0^2-\mathcal{S}_0}
\end{aligned}
\right.,\label{eq:fm2_up}
\\
   \mathcal{S}_1/t_0 & -\mathcal{S}_0>0\,, \left(\mathcal{S}_0/t_0-\mathcal{S}_{-1}^{(B)}\right)^2<\left(\mathcal{S}_0/t_0-\mathcal{S}_0^2/\mathcal{S}_1\right)^2\,,
    \label{fm2_B_condition_1}\\
    \mathcal{S}_2/t_0^2 & -\mathcal{S}_0>0\,, \left(\mathcal{S}_1/t_0^2-\mathcal{S}_{-1}^{(B)}\right)^2<\left(\mathcal{S}_1/t_0^2-\mathcal{S}_0^2/\mathcal{S}_1\right)^2 \,;
  \label{fm2_B_condition_2}
    \end{align}
\end{subequations}
where $\mathcal{S}_k = \{F_k,\mathcal{L}_k\}$ and the dependence on $s_0$ has been suppressed.
Eq.~\eqref{fm1_B_condition} is the required subsidiary condition to ensure the validity of \eqref{fm1_B}, as \eqref{fm2_B_condition_1} is to the first case of \eqref{eq:fm2_up}, and \eqref{fm2_B_condition_2} is to the second. 
The two upper bound expressions of $\mathcal{S}_{-2}$ in Eq.~\eqref{eq:fm2_up} are obtained through different methods and are numerically equivalent in the energy range of interest.
The lower bounds in \eqref{fm1_B} and \eqref{eq:fm2_up} ({\it i.e.}, $\mathcal{S}^\downarrow_k$, $k\in\{-1,-2\}$) are obtained from the general Cauchy-Schwarz  inequality 
\begin{equation}\label{eq:Cauchy}
    \frac{\mathcal{S}_k}{\mathcal{S}_{k+1}} \leq \frac{\mathcal{S}_{k-1}}{\mathcal{S}_{k}}\,,
\end{equation}
which is a special case of the general H\"older inequality developed for QCD sum rules in Refs.~\cite{Li:2024frm,Benmerrouche:1995qa,Ho:2018cat}.
As discussed below, Eq.~\eqref{eq:Cauchy} provides important constraints in our analysis for $k=1$. 

All $F_k$ and $\mathcal{L}_k$ with $k<0$ used in \eqref{eq:a_fit_exp} can be bounded using \eqref{eqs:fm1}--\eqref{eqs:fm2}. These bounds then produce upper and lower bound expressions for the $\tau$ anomalous magnetic moment, $\alb \leq \at^\mathrm{QCD} \leq \aub$, by substituting in Eq.~\eqref{eq:a_fit_exp}:

\begin{subequations}\label{eqs:a_sr_upper_and_lower}
    \begin{multline}
    \alb  =  \; 4\alpha^2 \biggl\{ 
    A_{-1} F_{-2}^\downarrow + A_{0}F_{-1}^\downarrow + \ga B_{1,0} \, \mathcal{L}_{-1}^\downarrow(\tau_1) + B_{1,1} \, \mathcal{L}_{0}(\tau_1) + B_{1,2} \, \mathcal{L}_{1}(\tau_1) + B_{1,3} \, \mathcal{L}_{2}(\tau_1) \dr \\
     + \ga B_{2,0} \, \mathcal{L}_{-1}^\downarrow(\tau_2) + B_{2,1} \, \mathcal{L}_{0}(\tau_2) + B_{2,2} \, \mathcal{L}_{1}(\tau_2) + B_{2,3} \, \mathcal{L}_{2}(\tau_2) \dr \biggr\},\label{eq:a_sr_explicit_lower}
    \end{multline}
    \begin{multline}
    \aub  =  \; 4\alpha^2 \biggl\{ 
    A_{-1} F_{-2}^\uparrow + A_{0}F_{-1}^\uparrow + \ga B_{1,0} \, \mathcal{L}_{-1}^\uparrow(\tau_1) + B_{1,1} \, \mathcal{L}_{0}(\tau_1) + B_{1,2} \, \mathcal{L}_{1}(\tau_1) + B_{1,3} \, \mathcal{L}_{2}(\tau_1) \dr \\
     + \ga B_{2,0} \, \mathcal{L}_{-1}^\uparrow(\tau_2) + B_{2,1} \, \mathcal{L}_{0}(\tau_2) + B_{2,2} \, \mathcal{L}_{1}(\tau_2) + B_{2,3} \, \mathcal{L}_{2}(\tau_2) \dr \biggr\},\label{eq:a_sr_explicit_upper}
    \end{multline}
\end{subequations}
where the $s_0$ dependence is suppressed. 

Alternative methods of fitting the kernel function have been explored. For instance, an approach which uses LSRs exclusively 
is discussed in  Refs.~\cite{Li:2024frm,Steele1991,Benmerrouche:1993rj},
which used over- and under-estimation of $K_\ell(t)$ in Eq.~\eqref{eq:exact-kernel} to bound $a_\ell^\mathrm{HVP,LO}$. However, both alternative fitting methods lead to bounds on $\auhvplo$ which are weaker than those  in Ref.~\cite{Li:2024frm}, which used only FESRs.
The combined expression using both FESRs and LSRs for $\at^{\rm QCD}$ in 
Eq.~\eqref{eq:a_fit_exp} 
aims to take advantage of the complementary strengths of both approaches: a polynomial kernel from the FESRs and an exponential kernel from the LSRs.

\section{QCD Inputs for Sum Rules}\label{sec:QCD_input}
To calculate the QCD correlation function $\Pi^H$ in the dispersive expression \eqref{eq:a_fit_exp}
for $\atQCD$, we use the light quark vector current $j_\mu=\bar{q}\gamma_\mu q$ [see Eq.~\eqref{tom_corr_fn}]. We consider the operator product expansion (OPE)\cite{Shifman:1978bx,Shifman:1978by,Reinders:1984gu} 
(see also Refs.~\cite{Reinders:1984sr,Pascual1984,Narison:2002woh,Steele1991}) of the vector-current correlator to five-loop order in perturbative contributions in the $\overline{\mathrm{MS}}$-scheme and chiral limit \cite{Baikov:2008jh,Gorishnii:1990vf,Surguladze:1990tg,Chetyrkin:1996ez,Chetyrkin:1979bj,Dine:1979qh,Celmaster:1979xr} (see also Refs.~\cite{Ahmady:2002pa,dEnterria:2022hzv}). We consider light-quark mass corrections to LO, and NLO corrections in the dimension-four QCD quark and gluon condensates, as well as LO corrections in dimension-six QCD quark condensates~\cite{Surguladze:1990sp,Wang:2016sdt,Chetyrkin:1985kn},
\begin{equation} \label{eq:correlator_sum}
    \Pi^H \left(Q^2\right)= \sum_{q} Q_q^2 \; \Pi_q\left(Q^2\right),
\end{equation}
\vspace{-0.6cm}
\begin{multline}
    \Pi_q\left(Q^2\right) =  \frac{1}{4\pi^2} \Pi^\mathrm{pert}_q\left(Q^2\right)
    - \frac{3 m_q^2(\nu)}{2\pi^2 Q^2} 
    + 2\langle m_q \bar{q}q\rangle\frac{1}{Q^4}\left(1+\frac{1}{3}\xi_q \frac{\alpha_s(\nu)}{\pi}\right)\\
    +\frac{1}{12\pi}\langle \alpha_s G^2 \rangle \frac{1}{Q^4}
    \left(1+\frac{7}{6}\xi_g\frac{\alpha_s(\nu)}{\pi}\right)
    -\frac{224}{81}\pi \alpha_s \langle \bar{q}\bar{q}qq\rangle \frac{1}{Q^6}\,,
    \label{eq:correlator}
\end{multline}
where $Q_q$ are the charges for light quarks with $q = \{u,\, d,\, s\}$,  $\xi_q$ and $\xi_g$ are renormalization group (RG) coefficients (listed in Table~\ref{tab:coefficients}) whose determinations are discussed in Sec.~\ref{sec:analysis}.
The imaginary part of the perturbative contribution \eqref{eq:correlator} is
\begin{gather}
  \frac{1}{\pi}\mathrm{Im}\Pi^\mathrm{pert}\left(t,\nu\right) = 1+\sum_{n=1}^{\infty}\ga \frac{\alpha_s(\nu)}{\pi} \dr^{n} \sum_{m=0}^{n-1} T_{n,m}\log^m\left(\frac{\nu^2}{t}\right)\,,
    \label{eq:pertseries}
\end{gather}
where the $\overline{\mathrm{MS}}$-scheme coefficients $T_{n,m}$ are given up to five-loop order in Table~\ref{tab:coefficients}. We tabulate $T_{n,m}$ for $n_f=3$ and $4$, corresponding to the number of active quark flavours over the renormalization-scale range relevant to our analysis.

\begin{table}[htbp]
\centering
\small
\renewcommand{\arraystretch}{1.4}
\begin{tabular}{c cccc c cccc}
\hline\hline
\multirow{2}{*}{$T_{n,m}$} & \multicolumn{4}{c}{$n_f = 4$} & & \multicolumn{4}{c}{$n_f = 3$} \\
\cline{2-5} \cline{7-10}
& $m=0$ & $m=1$ & $m=2$ & $m=3$ & & $m=0$ & $m=1$ & $m=2$ & $m=3$ \\
\hline
$n=1$ & $1$        & --        & --        & --                       & & $1$        & --        & --       & --      \\
$n=2$ & $1.52453$  & $25/12$   & --        & --                       & & $1.63982$  & $9/4$     & --       & --      \\
$n=3$ & $-11.6856$ & $9.56054$ & $625/144$ & --                       & & $-10.2839$ & $11.3792$ & $81/16$  & --      \\
$n=4$ & $-92.91$   & $-56.90$  & $36.56$   & $15625/1728$             & & $-106.896$ & $-46.2379$& $47.4048$& $729/64$\\
\hline
$\xi_q$ & \multicolumn{4}{c}{$31/25$} & & \multicolumn{4}{c}{$11/9$} \\
$\xi_g$ & \multicolumn{4}{c}{$-8/25$} & & \multicolumn{4}{c}{$-11/21$} \\
\hline\hline
\end{tabular}
\caption{$\overline{\mathrm{MS}}$-scheme coefficients 
$T_{n,m}$ for $n_f=4$ and $n_f=3$ (see \eg Refs~\cite{Li:2024frm,Baikov:2008jh,Ahmady:2002pa,vanRitbergen:1997va}), with the renormalization group (RG) coefficients $\xi_q$ and $\xi_g$.}
\label{tab:coefficients}
\end{table}

The FESRs and LSRs with $k=\{0,1,2\}$ can now be constructed using standard methodologies (see \eg Refs.~\cite{Floratos:1978jb,Hubschmid:1980rm,Bertlmann:1984ih,Caprini:1985ex,Shifman:1978bx,Shifman:1978by,Narison:2002woh})
from Eqs.~\eqref{eq:Fk_def} and \eqref{eq:Lk_def} using Eqs.~\eqref{eq:correlator}~and~\eqref{eq:pertseries}. Explicit expressions are provided in the Appendix. All the necessary QCD parameters are listed in Table~\ref{tab:parameters}. From these, one can calculate numerical values of upper and lower bounds for $\mathcal{S}_k (s_0)$ and $\at^\mathrm{QCD} (s_0)$ using Eqs.~\eqref{eqs:fm1}--\eqref{eqs:a_sr_upper_and_lower}.

\begin{table}[htb]
\centering
\renewcommand{\arraystretch}{1.4}
\begin{tabular}{ccc}
\hline\hline
\multicolumn{1}{c}{Parameter} & \multicolumn{1}{c}{Value} & \multicolumn{1}{c}{Source} \\ \hline
 $\alpha$ & $1/137.036$ & \cite{ParticleDataGroup:2026aaa} \\
 $\alpha_s\left(M_\tau\right)$ & $0.314\pm 0.014$ & \cite{ParticleDataGroup:2024cfk} \\
 $m_u(2\,\mathrm{GeV})$ & $2.16 \pm 0.07\,\mathrm{MeV}$ & \cite{ParticleDataGroup:2026aaa}\\
 $m_d(2\,\mathrm{GeV})$ & $4.70 \pm 0.07\,\mathrm{MeV}$ & \cite{ParticleDataGroup:2026aaa}\\
 $m_s(2\,\mathrm{GeV})$ & $92.9 \pm 0.7\gev$ & \cite{ParticleDataGroup:2026aaa} \\
 $f_\pi$ & $\left(0.13056\pm 0.00019\right)/\sqrt{2}$ GeV& \cite{ParticleDataGroup:2026aaa}\\
 $m_n \langle \bar{n}n\rangle$ & $-\frac{1}{2}f_\pi^2 m_\pi^2$ & \cite{Gell-Mann:1968hlm}  \\
 $m_s \langle \bar{s}s\rangle$  & $r_m r_c m_n \langle\overline{n}n\rangle $  & \cite{Harnett2021} \\
 $r_c \equiv \langle \bar{s}s\rangle/\langle \bar{n}n\rangle$ & $0.66 \pm 0.10 $  & \cite{Harnett2021} \\
 $m_s/m_n = r_m$ & $27.33^{+0.18}_{-0.14}$ & \cite{ParticleDataGroup:2026aaa}  \\
 $\langle \alpha G^2 \rangle$ & $\left(0.0649\pm 0.0035\right)\gev^4$ & \cite{Albuquerque:2023bex}\\
  $\kappa$ & $3.22\pm 0.5$ & \cite{Albuquerque:2023bex} \\
 $ \alpha_s \langle\bar{n}n\rangle^2$ & $\kappa\left(1.8\times 10^{-4}\right)\gev^6$  & \cite{Harnett2021} \\
  $ \alpha_s \langle \bar{s}s\rangle^2$ & $r_c^2 \alpha_s \langle \bar{n}n\rangle^2$ & \cite{Harnett2021}\\
\hline\hline
\end{tabular}
\caption{QCD parameters used in our calculations and uncertainty analyses, where $m_n = \left(m_u+m_d\right)/2$ and $\langle\bar{n}n\rangle = \langle\bar{u}u\rangle = \langle\bar{d}d\rangle$.}
\label{tab:parameters}
\end{table}

\newpage

\section{Analysis and Results}\label{sec:analysis}
From RG considerations, the renormalization scale is required to be $\nu = 1/\sqrt{\tau}$~\cite{Narison:1981ts} in LSRs, whereas $\nu = \sqrt{s_0}$ in FESRs (see \eg Refs.~\cite{Li:2024frm,Floratos:1978jb,Hubschmid:1980rm,Bertlmann:1984ih,Caprini:1985ex}).
As the dimension-four QCD condensate is evaluated up to NLO in the correlation function~\eqref{eq:correlator}, its RG behaviour must be treated consistently to the same order of correction. We reorganize the condensate contributions in terms of the RG-invariant combinations $\langle m_q \bar{q}q\rangle$ and $\langle \beta G^2\rangle+4\gamma\langle m_q \bar{q}q\rangle$~\cite{Pascual1984}, from which the RG coefficients $\xi_q$ and $\xi_g$ given in Table~\ref{tab:coefficients} are obtained. Both coefficients depend explicitly on $n_f$.

\begin{figure}
    \centering    \includegraphics[width=\linewidth]{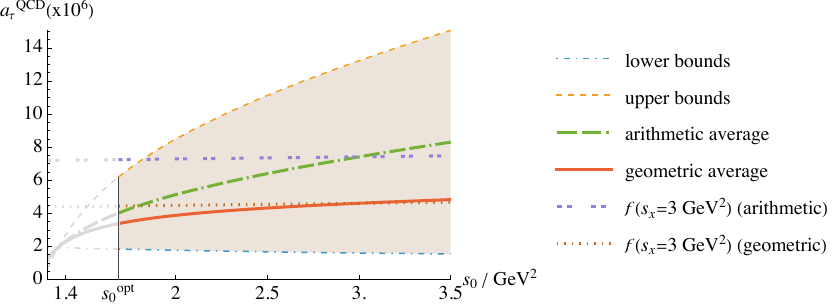}
    \caption{
    $\at^\mathrm{QCD}$ results including upper bound \eqref{eq:a_sr_explicit_upper}, lower bound \eqref{eq:a_sr_explicit_lower}, arithmetic and geometric averages of bounds, and the linear functions \eqref{eq:linear_fcn} at $s_x=3\gev^2$ for both averages used in the slope criterion. 
    The optimal $s_0^\mathrm{opt} = 1.689\gev^2$ is marked.
    The lighter shaded part of the results (where $s_0 < s^\mathrm{opt}_0$) is in the non-physical region that does not satisfy the inequality conditions.
    The shaded region (where $s_0 \geq s^\mathrm{opt}_0$) is the allowed range for $\at^\mathrm{QCD}$ using central values of the parameters in Table~\ref{tab:parameters}. 
    }
    \label{fig:results}
\end{figure}

In our methodology, we aim to find the optimal $s_0^\mathrm{opt}$ that provides us with the most restrictive bounds. At the same time, $s_0$ is constrained by the Cauchy-Schwarz inequality in Eq.~\eqref{eq:Cauchy}, which ensures the validity of hadronic spectral functions and hence $\mathcal{S}_k$. The value of $s_0$ is further constrained by all subsidiary conditions associated with our bound definitions \eqref{eqs:fm1}--\eqref{eqs:fm2}. 
Applying all inequality conditions \eqref{eqs:fm1}--\eqref{eq:Cauchy} to each $\mathcal{S}_k$ in Eq.~\eqref{eq:a_fit_exp} for $\at^\mathrm{QCD}$, we find the most stringent constraint on $s_0$ comes from the Cauchy-Schwarz inequality \eqref{eq:Cauchy} of 
LSRs with $\tau_2$, \ie $\mathcal{L}_1(\tau_2,s_0)/\mathcal{L}_0(\tau_2,s_0)<\mathcal{L}_2(\tau_2,s_0)/\mathcal{L}_1(\tau_2,s_0)$, leading to $s_0 \in \left[1.689,\infty\right)\gev^2$. 
During the bound analysis, we also find that the allowed band between $\aub(s_0)$ and $\alb(s_0)$ narrows down as $s_0$ decreases as shown in Fig.~\ref{fig:results}. Therefore, our lowest allowed value for $s_0$ gives the optimized (\ie strongest) 
sum-rule bounds on $\at^\mathrm{QCD}$.
We emphasize that it is the  $s_0$ dependence of the inequality bounds that is the crucial feature of our analysis, and the presence of bounds \eqref{eqs:fm1}--\eqref{eqs:fm2} on the (negative weight $k<0$) sum rules obscures the conventional interpretation of $s_0$ in a typical sum-rule analysis. 

To better understand Fig.~\ref{fig:results},
we observe that the upper and lower bounds approach each other as $s_0$ decreases and eventually cross below the valid $s_0$ region, suggesting a potential squeezed inequality prediction for $a_\tau^\mathrm{QCD}$ at the point where $\alb(s_0)=\aub(s_0)$. Although the value at which these bounds coincide cannot be directly extracted because it lies outside the region allowed by Eqs.~\eqref{eqs:fm1}--\eqref{eq:Cauchy}, this observation motivates us to seek a prediction within the valid region using the known bounds $\alb(s_0)$ and $\aub(s_0)$. Therefore, we consider arithmetic and geometric averages of the upper and lower bounds, then assess which provides a more appropriate estimate of the true value of
$a_\tau^\mathrm{QCD}$. This assessment is done through a slope criterion comparing the slopes of these two averaging functions with the slope of the dispersion relation \eqref{eq:a_tau_QCD} at LO of perturbation theory
\begin{equation}
    \frac{\dd}{\dd s_0} a_\tau^\mathrm{QCD}(s_x) \simeq \frac{\alpha^2}{s_0 \pi^2} \sum_q Q_q^2\, K_\tau(s_0)\biggr|_{s_0=s_x},
\label{eq:disp-reln-slope}
\end{equation}
where the choice of $s_x$ should avoid large resonance regions of the hadronic spectral function. 
For the slope criterion, the linear function that has the slope given by the RHS of \eqref{eq:disp-reln-slope} and intersects with the corresponding average at $s_x$ is given as
\begin{equation}\label{eq:linear_fcn}
    f(s_0) = (s_0-s_x) \frac{\dd}{\dd s_0}a_\tau^\mathrm{QCD}(s_x)-a_\tau^\mathrm{ave}(s_x),
\end{equation}
where $a^\mathrm{ave}_\tau$ is either arithmetic or geometric average of $a_\tau^\uparrow$ and $a_\tau^\downarrow$. As shown in Fig.~\ref{fig:results}, at the characteristic value $s_x=3\,\mathrm{GeV}^2$ the tangent to the geometric average lies closer to the linear function in Eq.~\eqref{eq:linear_fcn} than that of the arithmetic average.
Other choices of large $s_x$ (\eg in region $3\gev^2<s_x<5\gev^2$) lead to the same conclusion. 
This agreement motivates our use of the geometric average of the bounds to obtain a prediction for $a_\tau^\mathrm{QCD}$.

Due to the nature of the FESRs~\eqref{eq:Fk_def} and LSRs~\eqref{eq:Lk_def}, the search for the optimal $s_0$ involves scanning from high to low energy to find the optimal value that provides the smallest window of $\left[ \alb, \aub \right]$.
In practice, this means we need to evolve the renormalization scale from the reference values of $\alpha_s(M_\tau)$ and $m_q(2\gev)$, which are both determined in the $n_f=4$ region, to lower $s_0$ potentially in the $n_f=3$ region.  
The inequality boundary $s_0^\mathrm{opt}=1.689\,\gev^2$ and the renormalization scale $1/\sqrt{\tau_i}$ lie above the charm threshold, and so $n_f=4$ expressions for the RG functions and their associated virtual corrections within the $T_{n,m}$ perturbative coefficients are implemented for our benchmark analysis; no $n_f=3$ matching is required. The resulting central value  benchmark bounds and geometric average values are 
\begin{gather}
a^\downarrow_\tau =1.844 \times 10^{-6}\,,~
a_\tau^\uparrow = 6.214 \times 10^{-6}
\label{eq:a_bounds_central}
\\
    \at^\mathrm{QCD}\bigr|_\mathrm{geo,ave} =  3.385\times 10^{-6}\,,~
    s_0^\mathrm{opt} = 1.689 \gev^2.
    \label{eq:a_geo_ave_central}
\end{gather}

Nevertheless it is interesting to explore the methodological uncertainty of choosing $n_f=3$ (for RG evolution and virtual corrections) 
for comparison to the $n_f=4$  benchmark analysis.  In this case we use 
$\alpha_s(M_\tau)$ and $m_q(2\gev)$ to obtain $n_f=3$ RG-evolution boundary values at the charm threshold $\alpha_s(m_c)$ and $m_q(m_c)$, a process outlined in  more detail within 
Refs.~\cite{Ho:2016owu,Li:2025ljo,Steele:1998qls} (in principle the small effect of flavour threshold discontinuities \cite{Chetyrkin:1997un} is also involved).
The effect on the benchmark central values of Eq.~\eqref{eq:a_geo_ave_central} is quite small  ($\delta s_0^\mathrm{opt}\approx 0.01\gev^2$,  $\delta \at^\mathrm{QCD}\bigr|_\mathrm{geo,ave}\approx 0.03 \times 10^{-6}$) and will be included as one source of theoretical uncertainty.

We also study the sensitivity of our $\at^\mathrm{QCD}$ prediction on the QCD parameters listed in Table~\ref{tab:parameters}. Variations in the vacuum saturation parameter $\kappa$, the dimension-four gluon condensate $\left<\alpha_s G^2\right>$, and the strange-quark condensate parameter $r_c$ dominate the uncertainties in $s_0^\mathrm{opt}$, the $\at^\mathrm{QCD}$ bounds, and hence also our predicted value of $\atQCD$. 
Including all sources of uncertainty in quadrature, our final result for light-quark contributions extracted from the upper/lower bounds via the geometric average  (as motivated by the slope criterion \eqref{eq:disp-reln-slope} illustrated in Fig.~\ref{fig:results}) is 
\begin{equation}\label{eq:a_geo_ave}
    \at^\mathrm{QCD}\bigr|_\mathrm{geo,ave} = \ga 3.385\pm 0.048 \dr \times 10^{-6}
    \text{ with } 
    s_0^\mathrm{opt} = 1.689\pm 0.162 \gev^2.
\end{equation}
The geometric average is found to be less sensitive to variations in the QCD input parameters than the arithmetic average, providing supporting rationale for the geometric average  in addition to the criterion of Eq.~\eqref{eq:disp-reln-slope}.

\begin{figure}
    \centering
    \includegraphics[width=\linewidth]{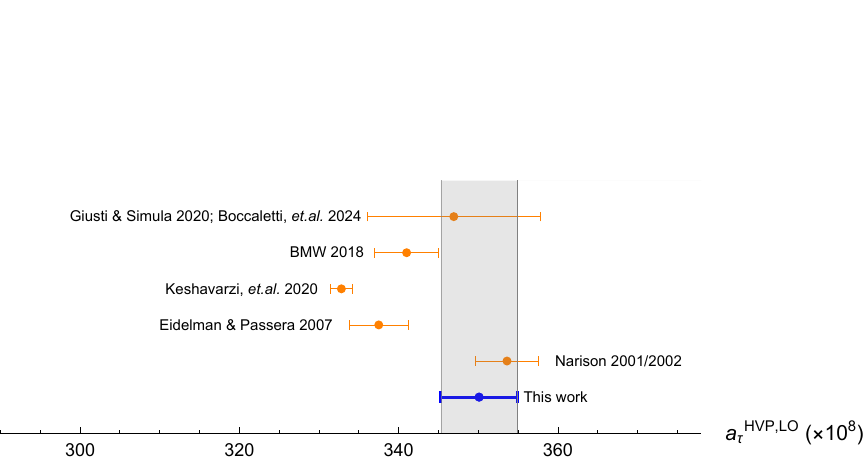}
    \caption{Summary of various determinations of $\athvplo$~\cite{Giusti:2020efo,Boccaletti:2024guq,Budapest-Marseille-Wuppertal:2017okr,Keshavarzi:2019abf,Eidelman:2007sb,Narison:2001jt} with uncertainties from Table~\ref{tab:lit_compare}. The grey band highlights our result~\eqref{eq:result_supplimented} based on the geometric average of our determined bounds.}
    \label{fig:compare}
\end{figure}

In order to compare our prediction with other evaluations (listed in Table~\ref{tab:lit_compare}), we supplement our light-quark result with resonance contributions above the charm threshold. Because our calculations do not explicitly include contributions from $c$ and $b$ quarks, which are expected to be sub-leading, we extract the charmonium and bottomonium resonance contributions from Ref.~\cite{Keshavarzi:2019abf} $
a_{\tau,\bar{c}c,\bar{b}b}^\mathrm{HVP,LO} = \ga 11.64\pm0.28 \dr \times 10^{-8}
$. Including these contributions, we obtain 
\begin{equation}\label{eq:result_supplimented}
    \athvplo = \ga 3.501\pm0.048\dr \times 10^{-6}.
\end{equation}
Our result \eqref{eq:result_supplimented} can then be compared with the Table~\ref{tab:lit_compare}
current data-driven value of
$(3.328\pm 0.014)\times 10^{-6}$~\cite{Keshavarzi:2019abf} 
and lattice QCD value of $(3.410\pm0.040)\times10^{-6}$ from Ref.~\cite{Budapest-Marseille-Wuppertal:2017okr} or $\ga3.469\pm0.109\dr\times10^{-6}$ from Refs.~\cite{Giusti:2020efo,Boccaletti:2024guq}. 
Fig.~\ref{fig:compare} compares our final value, Eq.~\eqref{eq:result_supplimented}, with a selection of other determinations of $\athvplo$ from Table~\ref{tab:lit_compare}. Our result is in tension with data-driven predictions of Refs.~\cite{Keshavarzi:2019abf,Eidelman:2007sb}, but is in better agreement with the LQCD results derived from Refs.~\cite{Giusti:2020efo,Boccaletti:2024guq} and \cite{Budapest-Marseille-Wuppertal:2017okr}, as well as with the determinations of Ref.~\cite{Narison:2001jt} that combines experimental inputs and QCD predictions.
Since our sum-rule method itself is based on the dispersion relation \eqref{eq:a_tau_0}, its agreement with other methods suggests a potential reconciliation of disagreement from different approaches. The fundamental sum-rule methodology can therefore help bridge the approaches and lead to a unified theoretical value of $a_\tau$.

As noted above, the combined expression for $\at^{\rm QCD}$ in Eq.~\eqref{eq:a_fit_exp} follows Ref.~\cite{Li:2024frm} by combining polynomial and exponential kernels from FESRs and LSRs, respectively.
This combined approach does not directly lend itself to an analysis of $\auhvplo$ because the muon kernel function  $K_\mu(t)$ (\ie Eq.~\eqref{eq:exact-kernel} with $m_\mu$ instead of $m_\tau$) is extremely well-approximated by a single polynomial $K_\mu(t)\approx m_\mu^2/(3t)$. However, 
Ref.~\cite{Giusti:2020efo} provides a LQCD ratio value of
\begin{equation}
R_{\mu/\tau} = \ga\frac{m_\tau}{m_\mu}\dr^2 \frac{\auhvplo}{\athvplo} = 5.83 (17)
\label{eq:tom_lattice_ratio}
\end{equation}
that can be combined with our  Eq.~\eqref{eq:result_supplimented} $\athvplo$ value to obtain the following combined QCD determination from lattice and sum-rule methods 
\begin{equation}
\auhvplo = (7.217\pm0.233)\times10^{-8}\,.
\label{eq_tom_a_mu}
\end{equation}
The result \eqref{eq_tom_a_mu} has significantly reduced theoretical uncertainty for $\auhvplo$  than the result obtained  in \cite{Li:2024frm}, providing additional information to help reconcile differences between theoretical and data-driven determinations of $\auhvplo$ (see \eg Refs.~\cite{
Leplumey:2025kvv,Colangelo:2018mtw,Colangelo:2022prz,Stoffer:2023gba,Davier:2019can,Keshavarzi:2019abf,Aoyama:2020ynm,Keshavarzi:2018mgv,
Davier:2023fpl,Castro:2024prg,
RBC:2018dos,Giusti:2019xct,Borsanyi:2020mff,Lehner:2020crt,Wang:2022lkq,Aubin:2022hgm,Ce:2022kxy,ExtendedTwistedMass:2022jpw,RBC:2023pvn,Kuberski:2024bcj,Boccaletti:2024guq,Spiegel:2024dec,RBC:2024fic,Djukanovic:2024cmq,ExtendedTwistedMass:2024nyi,MILC:2024ryz}), and yields a QCD sum-rule prediction comparable to the world theoretical average, $\auhvplo\bigr|_\mathrm{WP25} = 7.132(61)\times 10^{-8}$~\cite{Aguillard:2025fij}.

\begin{table}
    \centering
    \renewcommand{\arraystretch}{1.4}
    \begin{tabular}{c|c|c|c}
    \hline\hline
        LO Value ($\times10^8$) & NLO Value ($\times10^8$) & Method & Source\\\hline
        $\left[210,320\right]$& & Bounds from LSR (light quark) & \cite{Benmerrouche:1993rj}\\
        $360\pm40$& $-12\pm 2$ & Data-driven & \cite{Samuel:1990su}\\
        $320\pm10$& & Relativistic constituent quark model & \cite{Holdom:1993ad}\\
        $307\pm2$& & Operator Product Expansion & \cite{Hamzeh:1996np}\\
        $353.6\pm 4.0$ &$27.6\pm5.8$ & QCD + experimental input& \cite{Narison:2001jt}\\
        $337.5\pm3.7$& $7.6\pm2$& Data-driven & \cite{Eidelman:2007sb}\\
        $332.81\pm1.39$& & Data-driven & \cite{Keshavarzi:2019abf}\\
        $341.0\pm4.0$ & & Lattice QCD & \cite{Budapest-Marseille-Wuppertal:2017okr}\\
        $346.92\pm10.87$& &Lattice ratio $R_{\mu/\tau}$
        & \cite{Giusti:2020efo} \& \cite{Boccaletti:2024guq}\\
        $350.1 \pm 4.8$ & & Bounds from FESRs and LSRs & This work\\
        \hline\hline
    \end{tabular}
    \caption{Predictions of $a_\tau^\mathrm{HVP}$ from various theoretical approaches. Recent values are 
    summarized and plotted in Fig.~\ref{fig:compare}.
    }
    \label{tab:lit_compare}
\end{table}
\section{Summary}\label{sec:summary}
This work has presented a sum-rule analysis of bounds on the leading order hadronic vacuum polarization contributions to the anomalous magnetic moment of tau lepton, $\athvplo$. We model the QCD contributions $\atQCD$ with a polynomial combination of Laplace QCD sum rules (LSRs) and finite-energy QCD sum rules (FESRs) with different weights. The upper and lower bounds for $\at^\mathrm{QCD}$ are constructed from H\"older, Cauchy-Schwarz and related fundamental inequalities developed for FESRs in~\cite{Li:2024frm} and extended to LSRs in this work.

Our bounds on $\atQCD$ are calculated using the light-quark QCD contributions in perturbation theory up to five-loop order in the $\overline{\mathrm{MS}}$-scheme in the chiral limit, LO light-quark mass corrections, NLO dimension-four QCD condensates, and LO dimension-six QCD condensates. 
We develop a selection criterion to extract a determination from these bounds, motivating the prediction for light-quark contributions $\atQCD = \ga 3.385\pm 0.048 \dr \times 10^{-6}$ given in Eq.~\eqref{eq:a_geo_ave} obtained by taking a geometric average of the upper and lower bounds at an optimal sum-rule energy scale (see Fig.~\ref{fig:results}). This light-quark $\atQCD$ result is then supplemented with charmonium and bottomonium resonance contributions for our final determination of $\athvplo = \ga 3.501\pm0.048\dr \times 10^{-6}$ in Eq.~\eqref{eq:result_supplimented}. 
Additionally, a combined $\auhvplo$ QCD determination from lattice and sum-rule methods of
$\auhvplo=(7.217\pm0.233)\times10^{-8}$ given in Eq.~\eqref{eq_tom_a_mu} is then obtained
via Eqs.~\eqref{eq:result_supplimented} and \eqref{eq:tom_lattice_ratio},  providing new theoretical information to help reconcile differences between other determinations. 

We have developed a robust methodology for determining bounds in negative-weight QCD sum rules, carefully taking into account renormalization-group behaviour and flavour thresholds. 
Our final prediction for $\athvplo$
is comparable to values obtained from other approaches (see Table~\ref{tab:lit_compare} and Fig.~\ref{fig:compare}), providing additional theoretical insights on the determination of $\at$ and enabling comparison with future experimental measurements as their precision improves. 
The new methodologies  established in this work could be valuable  for  QCD sum-rule determinations of other low-energy observables.

\section{{Acknowledgments}}
T.G. Steele is grateful for research funding from the Natural Sciences \& Engineering Research Council of Canada (NSERC Grant SAPIN-2021-00024).  
S.~Li is supported by the Deutsche Forschungsgemeinschaft (DFG) 
through the Collaborative Research Center 1660 ``Hadrons and Nuclei as Discovery Tools'' (grant 514321794).

\section*{Appendix: Sum-rule Expressions}\label{sec:appendix}
For each light quark flavour $q$, the FESRs \eqref{eq:Fk_def} with $k = {0,1,2}$ are constructed using the standard methodology~\cite{Floratos:1978jb,Hubschmid:1980rm,Bertlmann:1984ih,Caprini:1985ex}. 
Using Eqs.~\eqref{eq:correlator} and \eqref{eq:pertseries}, we construct the following FESRs:
\begin{subequations}\label{eqs:FESR_explicit}
\begin{multline}
    F_0^q\left(s_0\right) =  \frac{1}{4\pi^2}\Bigg[1  + \frac{\alpha_s(\nu)}{\pi}T_{1,0} + \left(\frac{\alpha_s(\nu)}{\pi}\right)^2\left(T_{2,0}+T_{2,1}\right)+ \left(\frac{\alpha_s(\nu)}{\pi}\right)^3\left(T_{3,0}+T_{3,1}+2T_{3,2}\right)\\ 
    + \left(\frac{\alpha_s(\nu)}{\pi}\right)^4\left(T_{4,0}+T_{4,1}+2T_{4,2}+6T_{4,3}\right)\Bigg]s_0
    -\frac{3}{2\pi^2}  m_q(\nu)^2\,,
    \label{eq:fesr-k-0}
\end{multline}
\begin{align}
 F_1^q\left(s_0\right) =  \frac{1}{8\pi^2}\Bigg[1 + \frac{\alpha_s(\nu)}{\pi}T_{1,0} & + \left(\frac{\alpha_s(\nu)}{\pi}\right)^2\left(T_{2,0}+\frac{1}{2}T_{2,1}\right)+ \left(\frac{\alpha_s(\nu)}{\pi}\right)^3\left(T_{3,0}+\frac{1}{2}T_{3,1}+\frac{1}{2}T_{3,2}\right) \nonumber\\ 
    & + \left(\frac{\alpha_s(\nu)}{\pi}\right)^4\left(T_{4,0}+\frac{1}{2}T_{4,1}+\frac{1}{2}T_{4,2}+\frac{3}{4}T_{4,3}\right)\Bigg]s_0^2 
    \nonumber\\
    &
    - 2\langle m_q \bar{q}q\rangle\left(1+\frac{1}{3}\xi_q\frac{\alpha_s(\nu)}{\pi}\right)
    - \frac{1}{12\pi}\langle \alpha_s G^2 \rangle\left(1+\frac{7}{6}\xi_g\frac{\alpha_s(\nu)}{\pi}\right)\,,
    \label{eq:fesr-k-1}
\end{align}
\begin{multline}
    F_2^q\left(s_0\right) =  \frac{1}{12\pi^2}\Bigg[1 + \frac{\alpha_s(\nu)}{\pi}T_{1,0} + \left(\frac{\alpha_s(\nu)}{\pi}\right)^2\left(T_{2,0}+\frac{1}{3}T_{2,1}\right)+ \left(\frac{\alpha_s(\nu)}{\pi}\right)^3\left(T_{3,0}+\frac{1}{3}T_{3,1}+\frac{2}{9}T_{3,2}\right)\\ 
    + \left(\frac{\alpha_s(\nu)}{\pi}\right)^4\left(T_{4,0}+\frac{1}{3}T_{4,1}+\frac{2}{9}T_{4,2}+\frac{2}{9}T_{4,3}\right)\Bigg]s_0^3 
    - \frac{224}{81} \pi \alpha_s \langle \bar{q}\bar{q}qq\rangle  \,,
    \label{eq:fesr-k-2}
\end{multline}
\end{subequations}
where the renormalization scale is fixed to $\nu = \sqrt{s_0}$
(see Refs.~\cite{Floratos:1978jb,Hubschmid:1980rm,Bertlmann:1984ih,Caprini:1985ex,Li:2024frm}). 

Laplace QCD sum rules (LSRs) \eqref{eq:Lk_def} are constructed using standard methods~\cite{Shifman:1978by,Shifman:1978bx,Narison:2002woh} from
Eqs.~\eqref{eq:correlator} and \eqref{eq:pertseries} giving
\begin{subequations}\label{eqs:LSR_explicit}
\begin{multline}\label{L0}
    L_0^q\left(\tau,s_0 \right)=\frac{1}{4\pi^2\tau}\left[ 
     f_{0,0}\left(\tau s_0\right)  +
    \sum_{k=0}^3 f_{0,k}\left(\tau s_0\right) \sum_{j=k+1}^4 T_{j,k} \left( \frac{\alpha_s(\nu)}{\pi}\right)^j
    \right]-\frac{3}{2\pi^2}m_q(\nu)^2   
    \\
    + 2\langle m_q \bar{q}q\rangle\left(1+\frac{1}{3}\xi_q\frac{\alpha_s(\nu)}{\pi}\right)\tau
   + \frac{1}{12\pi}\langle \alpha_s G^2 \rangle\left(1+\frac{7}{6}\xi_g\frac{\alpha_s(\nu)}{\pi}\right)\tau
   - \frac{112}{81} \pi \alpha_s \langle \bar{q}\bar{q}qq\rangle\tau^2\,,
\end{multline}
\begin{multline}\label{L1}
    L_1^q\left(\tau,s_0\right)=\frac{1}{4\pi^2\tau^2}\left[ 
     f_{1,0}\left(\tau s_0\right)  +
    \sum_{k=0}^3 f_{1,k}\left(\tau s_0\right) \sum_{j=k+1}^4 T_{j,k} \left( \frac{\alpha_s(\nu)}{\pi}\right)^j
    \right]    
   \\
    - 2\langle m_q \bar{q}q\rangle\left(1+\frac{1}{3}\xi_q\frac{\alpha_s(\nu)}{\pi}\right)
   - \frac{1}{12\pi}\langle \alpha_s G^2 \rangle\left(1+\frac{7}{6}\xi_g\frac{\alpha_s(\nu)}{\pi}\right)
   + \frac{224}{81} \pi \alpha_s \langle \bar{q}\bar{q}qq\rangle\tau\,,
\end{multline} 
\begin{equation}\label{L2} 
    L_2^q\left(\tau,s_0\right)=\frac{1}{4\pi^2\tau^3}\left[ 
     f_{2,0}\left(\tau s_0\right)  +
    \sum_{k=0}^3 f_{2,k}\left(\tau s_0\right) \sum_{j=k+1}^4 T_{j,k} \left( \frac{\alpha_s(\nu)}{\pi}\right)^j
    \right]
   - \frac{224}{81} \pi \alpha_s \langle \bar{q}\bar{q}qq\rangle\,, 
\end{equation}
where\\
\begin{equation}\label{f_jk_eq}
f_{j,k}\left(\tau s_0\right)=\int_0^{\tau s_0} z^j\left[\log{\left(\frac{1}{z}\right)} \right]^k e^{-z} \,dz\,.
\end{equation} 
\end{subequations}
The renormalization scale in the LSRs \eqref{L0}--\eqref{f_jk_eq} is set to $\nu = 1/\sqrt{\tau}$ 
~\cite{Narison:1981ts}.

One should keep in mind that the sum rules used in \eqref{eqs:a_sr_upper_and_lower} for $\athvplo$ bounds determinations and Eqs.~\eqref{eqs:fm1}--\eqref{eqs:fm2} for $s_0$ analyses are sums of all light-quark FESRs \eqref{eqs:FESR_explicit} and LSRs \eqref{eqs:LSR_explicit} combined with quark charge prefactors as defined in 
Eq.~\eqref{eq:correlator_sum}:
\begin{equation}
    \mathcal{S}_k (s_0) = \sum_q Q_q^2\, \mathcal{S}_k^q (s_0),
\end{equation}
where $\mathcal{S}_k^q = \{F_k^q,\mathcal{L}_k^q\}$, $Q_u^2 = 4/9$, $Q_d^2 = Q_s^2 = 1/9$, and $k\in\mathbb{Z}$.

\bibliographystyle{h-physrev}
\bibliography{g-2.bib}

\end{document}